\documentclass[twocolumn,amsmath,amssymb,floatfix]{revtex4}

\usepackage{graphicx}
\usepackage{dcolumn}
\usepackage{bm}

\graphicspath{{Figures/}} \setkeys{Gin}{width=\linewidth}

\begin{document}

\title{Aharonov-Bohm oscillations in p-type GaAs quantum rings}

\author{Boris Grbi\'{c}$^{*}$, Renaud Leturcq$^{*}$, Thomas Ihn$^{*}$,
Klaus Ensslin$^{*}$, Dirk Reuter $^{+}$, and Andreas D.
Wieck$^{+}$}

\affiliation{$^{*}$Solid State Physics Laboratory, ETH Zurich,
  8093 Zurich, Switzerland \\$^{+}$Angewandte Festk\"{o}rperphysik,
Ruhr-Universit\"{a}t Bochum, 44780 Bochum, Germany}

\begin{abstract}
We have explored phase coherent transport of holes in two p-type
GaAs quantum rings with orbital radii 420 nm and 160 nm fabricated
with AFM oxidation lithography. Highly visible Aharonov-Bohm (AB)
oscillations are measured in both rings, with an amplitude of the
oscillations exceeding $10\%$ of the total resistance in the case
of the ring with a radius of 160 nm. Beside the h/e oscillations,
we resolve the contributions from higher harmonics of the AB
oscillations. The observation of a local resistance minimum at B=0
T in both rings is a signature of the destructive interference of
the holes' spins. We show that this minimum is related to the
minimum in the h/2e oscillations.
\end{abstract}

\maketitle

The Aharonov-Bohm (AB) phase \cite{Aharonov59}, represents the
geometric phase acquired by the orbital wave function of the
charged particle encircling a magnetic flux line. This phase is
experimentally well  established and manifests itself through
oscillations in the resistance of mesoscopic rings as a function
of an external magnetic field. The spin part of the particle's
wave-function can acquire an additional geometric phase in the
systems with strong spin-orbit (SO) interactions
\cite{Engel00,Aronov93}. The SO induced phase additionally
modulates the resistance oscillations in mesoscopic rings. This SO
induced phase in solid-state systems has been recently the subject
of a number of experimental investigations
\cite{Morpugo98,Yau02,Yang04,Konig06,Bergsten06,Habib07,Grbic07}.

SO interactions are particularly strong in p-type GaAs
heterostructures, due to the p-like symmetry of the states at the
top of the valence band and the high effective mass of holes. The
presence of exceptionally strong SO interactions in carbon doped
GaAs heterostructure used for fabrication of rings investigated in
this work, is demonstrated by the simultaneous observation of the
beating in Shubnikov-de Haas (SdH) oscillations and weak
anti-localization in the measured magnetoresistance. The hole
density in an unpatterned sample is 3.8$\times$10$^{11}$ cm$^{-2}$
and the mobility is 200 000 cm$^2$/Vs at a temperature of 60 mK.
The strength of the Rashba spin-orbit interaction is estimated to
be $\Delta_{SO}\approx0.8$ meV, while the Fermi energy in the
system is $E_F=2.5$ meV.

Here we study AB oscillations in two quantum rings with radii 420
nm and 160 nm realized in this 2DHG with strong SO interactions.
In contrast to previous experiments on p-type GaAs rings, where
the signature of the phase acquired by the hole's spin was
attributed to the splitting of the h/e peak in the Fourier
spectrum \cite{Morpugo98,Yau02}, we have recently reported the
direct observation of beating in the measured resistance of the
quantum ring with an orbital radius of  420 nm \cite{Grbic07}. An
example of the observed beating in the gate configuration
V$_{pg1}=-172$ mV and V$_{pg2}=-188$ mV of the large ring is shown
in Fig. 1(a), while the corresponding splitting of the h/e Fourier
peak is shown in Fig. 1(c). In addition, we observe a resistance
minimum at $B=0$ T in all gate configurations, and attribute its
origin to the destructive interference of the hole spins
propagating along time reversed paths \cite{Grbic07}.

\begin{figure}[h]
\begin{center}\leavevmode
\includegraphics[width=0.4\textwidth]{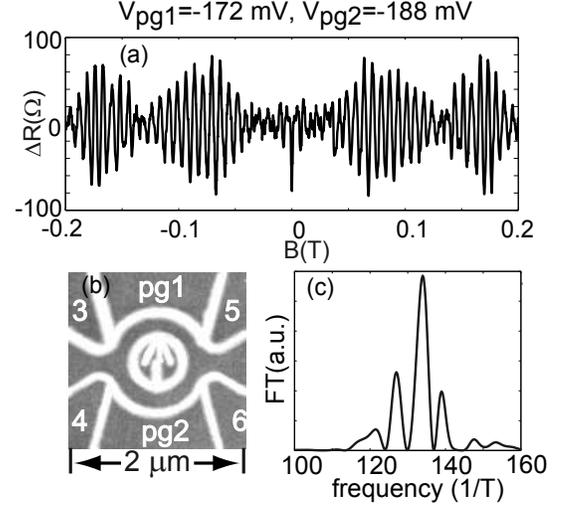}
\caption{(a) AB oscillations in the gate configuration
V$_{pg1}=-172$ mV and V$_{pg2}=-188$ mV obtained after subtraction
of the low-frequency background from the raw data. A clear beating
pattern is revealed in the AB oscillations. (b) AFM micrograph of
the large ring with an orbital radius of  420 nm  with
designations of the in-plane gates. (c) Splitting of the h/e
Fourier peak.} \label{figurename}
\end{center}
\end{figure}

We now focus on the magnetotransport measurements performed on the
smaller ring with an orbital radius of 160 nm (Fig. 2(a)). Fig.
2(b) shows the magnetoresistance of the small ring (oscillating
curve, blue online), together with a low-frequency background
resistance composed of the low frequency Fourier components of the
signal (smooth curve, red online) in the plunger gate
configuration V$_{pg1}=-78.5$ mV, V$_{pg2}=-222$ mV. AB
oscillations, obtained after subtraction of the low-frequency
background from the raw data, with a peak-to-peak amplitude of
$\sim4$ k$\Omega$  are clearly resolved (Fig. 2(c)). Therefore the
visibility of the AB oscillations is larger than $10\%$. The
Fourier transform of these oscillations reveals a split h/e peak,
with side peaks at 12 T$^{-1}$ and 17 T$^{-1}$. Besides, h/2e and
h/3e peaks are clearly visible (Fig. 2(d)). The electronic radius
of the holes' orbit of 150 nm, obtained from the period of the
oscillations of around 60 mT, agrees well with a lithographic mean
radius of the holes' orbit.

Due to the large period of the AB oscillations, only up to 10
oscillations are present in the magnetic field range (-0.3 T, +0.3
T) where SdH oscillations do not obscure the data analysis and no
beating can be seen in the raw data. Therefore, although the
amplitude of the AB oscillations in the case of the small ring is
quite large, a detailed analysis of the beating of the AB
oscillations can not be performed for this sample as it was done
in the case of the larger ring \cite{Grbic07}.

\begin{figure}[h]
\begin{center}\leavevmode
\includegraphics[width=0.4\textwidth]{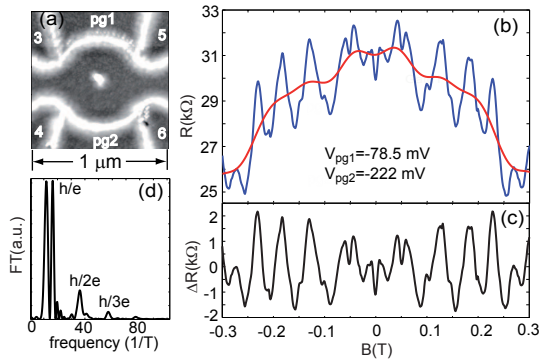}
\caption{(a) AFM micrograph of the small ring with an orbital
radius of 160 nm with designation of the gates. (b) Measured
magnetoresistance of the small ring (oscillating curve, blue
online) together with the low-frequency background resistance
(smooth curve, red online) in the plunger gate configuration
V$_{pg1}=-78.5$ mV, V$_{pg2}=-222$ mV. (c) AB oscillations
obtained after subtraction of the low-frequency background from
the raw data. (d) Fourier transform of the data with split h/e
peak and well defined h/2e and h/3e peaks.} \label{figurename}
\end{center}
\end{figure}

We further analyze the dependence of the AB oscillations on
plunger gates voltages. The evolution of the AB oscillations along
the line $V_{pg1}=0.5 \cdot V_{pg2}-20$mV in parameter space
$(V_{pg1}, V_{pg2})$ is investigated in Fig. 3. Fig. 3(a) shows
the raw data, while Fig. 3(b) and 3(c) show filtered h/e and h/2e
oscillations, respectively. When $V_{pg1}$ and $V_{pg2}$ increase
along the given line, both ring arms become narrower and the
holes' orbit inside the ring shrinks, causing the resistance of
the ring to increase continuously from $20-50$ k$\Omega$.

\begin{figure}[h]
\begin{center}\leavevmode
\includegraphics[width=0.3\textwidth]{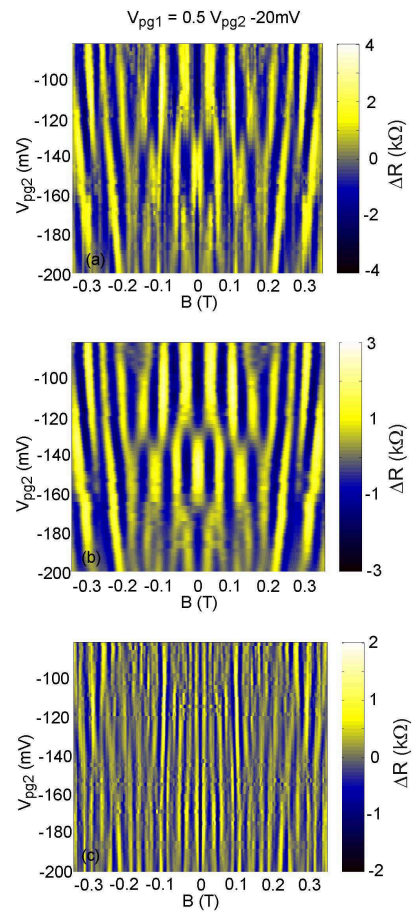}
\caption{ (a) Evolution of the AB
    oscillations upon changing plunger gate voltages along the line $V_{pg1}=0.5 \cdot V_{pg2}-20$mV in parameter space. (b)
Filtered h/e oscillations as a function of plunger gate voltages,
showing phase jumps. (c) Filtered h/2e oscillations as a function
of plunger gate voltages, showing the minimum at $B=0$ T at all
gate voltages.} \label{figurename}
\end{center}
\end{figure}

We again observe a resistance minimum at $B=0$T in all gate
configurations (Fig. 3(a)) and find that it is related to the
minimum in the h/2e oscillations at $B=0$T (Fig. 3(c)), consistent
with the results from the large ring sample
\cite{Grbic07,GrbicThesis07}. It should be emphasized that the
observed minimum is not caused by weak-antilocalization in the
ring leads, since the weak-antilocalization dip in bulk
two-dimensional samples has a much smaller magnitude (less than
$1\Omega$) than the minimum at $B=0$T in the rings
\cite{GrbicThesis07}. The resistance minimum at $B=0$T is a result
of the destructive interference of the holes' spins in the ring.

The presence of phase jumps in the h/e oscillations (Fig. 3(b))
and their absence in the h/2e oscillations (Fig. 3(c)) is also
observed. The fact that the phase of the AB oscillations can not
change continuously, but only in discrete steps of $\pi$ is a
consequence of the Onsager relations $G_{ij}(B)=G_{ji}(-B)$. For
changes of the plunger gates along the line $V_{pg1}=0.5 \cdot
V_{pg2}-20$mV we indeed observe phase jumps of $\pi$ in h/e
oscillations at lower magnetic fields up to 0.2 T, but at the
higher fields, above 0.2T, we find continuous monotonic shifts of
the AB minima and maxima (Fig. 3(a) and 3(b)). We attribute these
continuous shifts of the AB minima and maxima towards higher
fields upon increasing plunger gate voltages $V_{pg1}$ and
$V_{pg2}$ to an increase of the AB oscillation frequency upon
continuous shrinking of the holes' orbit within the ring.
Continuous, but non-monotonic top-gate induced shifts of the AB
minima and maxima were recently observed in HgTe quantum rings
\cite{Konig06}, and this behavior is interpreted as a
manifestation of the Aharonov-Casher phase.

In conclusion, we have measured highly visible Aharonov-Bohm
oscillations in two quantum rings with radii 420 nm and 160 nm
fabricated by AFM oxidation lithography on p-type GaAs
heterostructure with strong spin-orbit interaction. The visibility
of the AB oscillations exceeds $3\%$ in the larger ring and $10\%$
in the smaller ring. Beside the h/e oscillations, the higher
harmonics of the AB oscillations are resolved in both rings. A
resistance minimum at $B=0$T, present in both rings, points to the
signature of destructive interference of the holes' spins
propagating along time-reversed paths.






\end{document}